# A description of pseudorapidity distributions of charged particles produced in Au+Au collisions at RHIC energies

Z. J. Jiang[*], Dongfang Xu, and Yan Huang

*College of Science, University of Shanghai for Science and Technology, Shanghai 200093, China*

*jzj265@163.com

The charged particles produced in heavy ion collisions consist of two parts: One is from the freeze-out of hot and dense matter formed in collisions. The other is from the leading particles. In this paper, the hot and dense matter is assumed to expand according to the hydrodynamic model including phase transition and decouples into particles *via* the prescription of Cooper-Frye. The leading particles are as usual supposed to have Gaussian rapidity distributions with the number equaling that of participants. The investigations of this paper show that, unlike low energy situations, the leading particles are essential in describing the pseudorapidity distributions of charged particles produced in high energy heavy ion collisions. This might be due to the different transparencies of nuclei at different energies.

## 1. Introduction

The BNL Relativistic Heavy Ion Collider (RHIC) accelerates nuclei up to the center-of-mass energies from a dozen GeV to 200 GeV per nucleon. In the past decade, the measurements from such collisions have triggered an extensive research for the properties of matter at extreme conditions of very high temperature and energy densities [1-33]. One of the most important achievements from such research is the discovery that the matter created in nucleus-nucleus collisions at RHIC energies is in the state of strongly coupled quark-gluon plasma (sQGP) exhibiting a clear collective behavior nearly like a perfect fluid with very low viscosity [10-33].

The best approach for describing the space-time evolution of fluid-like sQGP is the relativistic hydrodynamics, which was first put forward by L. D. Landau in his pioneering work in 1953 [34].

However, since the partial differential equations of relativistic hydrodynamics are highly nonlinear, it is a formidable task to solve them analytically. This is the reason why, from the time of Landau until now, the exact solutions of relativistic hydrodynamics are mainly limited to 1+1 expansion for a perfect fluid with simple equation of state. To solve the equations of high-dimensional expansions especially for situations incorporating the effect of viscosities or pressure anisotropies, one has to resort to the means of computer simulations.

One of the most important applications of 1+1 dimensional hydrodynamics is the analysis of the pseudorapidity distributions of charged particles in high energy physics. In this paper, combing the effect of leading particles, we will discuss such distributions in the framework of hydrodynamic model including phase transition [10]. In section 2, a brief introduction is given to the theoretical model, presenting its exact solutions. The solutions are then used in section 3 to formulate the pseudorapidity distributions of charged particles resulted from the freeze-out of sQGP. Together with the contribution from leading particles, the results are then compared with the experimental observations performed by PHOBOS Collaboration at RHIC in Au+Au collisions at $\sqrt{s_{NN}}$ =200 and 19.6 GeV [5], respectively. The last section 4 is traditionally about conclusions.

## 2. A brief introduction to the model

Here, for the purpose of completeness and applications, we will list the key ingredients of the hydrodynamic model [10].

(1) The movement of fluid follows the continuity equation

$$\frac{\partial T^{\mu\nu}}{\partial x^{\nu}} = 0, \quad \mu, \nu = 0, 1, \tag{1}$$

where $x^{\nu} = (x^0, x^1) = (t, z)$, $t$ is the time and $z$ is the longitudinal coordinate along beam direction. $T^{\mu\nu}$ is the energy-momentum tensor, which, for a perfect fluid, takes the form

$$T^{\mu\nu} = (\varepsilon + p)u^{\mu}u^{\nu} - pg^{\mu\nu}, \tag{2}$$

where $g^{\mu\nu} = g_{\mu\nu} = \text{diag}(1,-1)$ is the metric tensor.

$$u^{\mu} = (u^0, u^1) = (\cosh y_F, \sinh y_F), \quad u^{\mu}u_{\mu} = 1, \tag{3}$$

is the 4-velocity of fluid, $y_F$ is its rapidity. $\varepsilon$ and $p$ in Eq. (2) are the energy density and

pressure of fluid, which meet the thermodynamical relations

$$\varepsilon + p = Ts, \ \mathrm{d}\varepsilon = T\mathrm{d}s, \ \mathrm{d}p = s\mathrm{d}T, \tag{4}$$

where $T$ and $s$ are the temperature and entropy density of fluid, respectively. To close Eq. (1), another relation, namely the equation of state

$$\frac{\mathrm{d}p}{\mathrm{d}\varepsilon} = \frac{s\mathrm{d}T}{T\mathrm{d}s} = c_s^2 \tag{5}$$

is needed, where $c_s$ is the sound speed of fluid, which takes different values in sQGP and in hadronic phase.

(2) Project Eq. (1) to the direction of $u_\mu$ and the direction perpendicular to $u_\mu$, respectively. This leads to equations

$$\frac{\partial(su^\nu)}{\partial x^\nu} = 0, \tag{6}$$

$$\frac{\partial(T\sinh y_F)}{\partial t} + \frac{\partial(T\cosh y_F)}{\partial z} = 0. \tag{7}$$

Eq. (6) is the continuity equation for entropy conservation. Eq. (7) means the existence of a scalar function $\phi$ satisfying relations

$$\frac{\partial \phi}{\partial t} = T\cosh y_F, \ \frac{\partial \phi}{\partial z} = -T\sinh y_F. \tag{8}$$

From $\phi$ and Legendre transformation, Khalatnikov potential $\chi$ is introduced *via* relation

$$\chi = \phi - tT\cosh y_F + zT\sinh y_F. \tag{9}$$

In terms of $\chi$, the variables $t$ and $z$ can be expressed as

$$\begin{aligned} t &= \frac{e^\theta}{T_0}\left(\frac{\partial \chi}{\partial \theta}\cosh y_F + \frac{\partial \chi}{\partial y_F}\sinh y_F\right), \\ z &= \frac{e^\theta}{T_0}\left(\frac{\partial \chi}{\partial \theta}\sinh y_F + \frac{\partial \chi}{\partial y_F}\cosh y_F\right), \end{aligned} \tag{10}$$

where $T_0$ is the initial temperature of fluid and $\theta = \ln(T_0/T)$. Through above equations, the coordinate base of $(t, z)$ is transformed to that of $(\theta, y_F)$, and Eq. (6) is translated into the so called telegraphy equation

$$\frac{\partial^2 \chi}{\partial \theta^2} - 2\beta \frac{\partial \chi}{\partial \theta} - \frac{1}{c_s^2} \frac{\partial^2 \chi}{\partial y_F^2} = 0, \ \beta = \frac{1-c_s^2}{2c_s^2}. \tag{11}$$

(3) Along with the expansions of matter created in collisions, it becomes cooler and cooler. As its temperature drops from the initial $T_0$ to the critical $T_c$, phase transition occurs. The matter transforms from sQGP state to hadronic state. The produced hadrons are initially in the violent and frequent collisions. The major part of these collisions is inelastic. Hence, the abundances of identified hadrons are in changing. Furthermore, the mean free paths of these primary hadrons are very short. The movement of them is still like that of a fluid meeting Eq. (11) with only difference being the value of $c_s$. In sQGP, $c_s = c_0 = 1/\sqrt{3}$, which is the sound speed of a massless perfect fluid, being the maximum of $c_s$. In the hadronic state, $c_s = c_h < c_0$. At the point of phase transition, that is as $T = T_c$, $c_s$ is discontinuous.

(4) The solution of Eq. (11) for the sector of sQGP is [10]

$$\chi_0(\theta, y_F) = \frac{q_0 c_0}{2} e^{\beta_0 \theta} I_0\left(\beta_0 c_0 \sqrt{y_0^2(\theta) - y_F^2}\right), \tag{12}$$

where $q_0$ is a constant determined by tuning the theoretical results to experimental data. $I_0$ is the 0th order modified Bessel function of the first kind, and

$$\beta_0 = \frac{1-c_0^2}{2c_0^2} = 1, \ y_0(\theta) = \frac{\theta}{c_0}. \tag{13}$$

In the sector of hadrons, the solution of Eq. (11) is [10]

$$\chi_h(\theta, y_F) = \frac{q_0 c_0}{2} B(\theta) I_0[\lambda(\theta, y_F)], \tag{14}$$

where

$$B(\theta) = e^{\beta_h(\theta - \theta_c) + \beta_0 \theta_c}, \ \lambda(\theta, y_F) = \beta_h c_h \sqrt{y_h^2(\theta) - y_F^2},$$
$$\beta_h = \frac{1-c_h^2}{2c_h^2}, \ y_h(\theta) = \frac{\theta - \theta_c}{c_h} + \frac{\theta_c}{c_0}, \ \theta_c = \ln\left(\frac{T_0}{T_c}\right). \tag{15}$$

## 3. The pseudorapidity distributions of charged particles

(1) The invariant multiplicity distributions of charged particles frozen out from sQGP

From Khalatnikov potential $\chi$, the rapidity distributions of charged particles frozen out from fluid-like sQGP read [35]

$$\frac{dN_{sQGP}}{dy_F} = \frac{q_0 c_0}{2} A(b) \left( \cosh y \frac{dz}{dy_F} - \sinh y \frac{dt}{dy_F} \right), \tag{16}$$

where $A(b)$ is the area of overlap region of collisions, being the function of impact parameter $b$ or centrality cuts. Inserting Eq. (10) into above equation, the part in the round brackets becomes

$$\cosh y \frac{dz}{dy_F} - \sinh y \frac{dt}{dy_F}$$
$$= \frac{1}{T} c^2 \frac{\partial}{\partial \theta} \left( \chi + \frac{\partial \chi}{\partial \theta} \right) \cosh(y - y_F) - \frac{1}{T} \frac{\partial}{\partial y_F} \left( \chi + \frac{\partial \chi}{\partial \theta} \right) \sinh(y - y_F). \tag{17}$$

With the expansions of hadronic matter, it continues becoming cooler. According to the prescription of Cooper-Frye [35], as the temperature drops to the freeze-out temperature $T_{FO}$, the inelastic collisions among hadrons cease. The yields of identified hadrons maintain unchanged becoming the measured results in experiments. The invariant multiplicity distributions of charged particles equal [10, 15, 35]

$$\frac{d^2 N_{sQGP}}{2\pi p_T dy dp_T} = \frac{1}{(2\pi)^3} \int \frac{dN_{sQGP}}{dy_F} \frac{m_T \cosh(y - y_F)}{\exp\{[m_T \cosh(y - y_F) - \mu_B]/T\} + \delta} \bigg|_{T=T_{FO}} dy_F, \tag{18}$$

where $m_T = \sqrt{m^2 + p_T^2}$ is the transverse mass of produced charged particle with rest mass $m$. $\mu_B$ in Eq. (18) is the baryochemical potential. For Fermi charged particles, $\delta = 1$ in the denominator of Eq. (18), and for Bosons, $\delta = -1$. The meaning of Eq. (18) is evident. It is the convolution of $dN_{sQGP}/dy_F$ with the energy of the charged particles in the state with temperature $T$.

The integral interval of $y_F$ in Eq. (18) is $[-y_h(\theta_f), y_h(\theta_f)]$. The integrand is evaluated with $T = T_{FO}$. At this moment, the fluid freezes out into the charged particles. Replacing $\chi$ in Eq. (17) by $\chi_h$ of Eq. (14), it becomes

$$\left( \cosh y \frac{dz}{dy_F} - \sinh y \frac{dt}{dy_F} \right) \bigg|_{T=T_{FO}}$$
$$= \frac{1}{T_{FO}} (\beta_h c_h)^2 B(\theta_{FO}) \left[ S(\theta_{FO}, y_F) \sinh(y - y_F) + C(\theta_{FO}, y_F) \cosh(y - y_F) \right], \tag{19}$$

where

$$S(\theta_{FO}, y_F) = \frac{\beta_h y_F}{\lambda(\theta_{FO}, y_F)} \left\{ \frac{\beta_h c_h y_h(\theta_{FO})}{\lambda(\theta_{FO}, y_F)} I_0[\lambda(\theta_{FO}, y_F)] \right.$$

$$\left. + \left[ \frac{\beta_h + 1}{\beta_h} - \frac{2\beta_h c_h y_h(\theta_{FO})}{\lambda^2(\theta_{FO}, y_F)} \right] I_1[\lambda(\theta_{FO}, y_F)] \right\}, \qquad (20)$$

$$C(\theta_{FO}, y_F) = \left\{ \frac{\beta_h + 1}{\beta_h} + \frac{[\beta_h c_h y_h(\theta_{FO})]^2}{\lambda^2(\theta_{FO}, y_F)} \right\} I_0[\lambda(\theta_{FO}, y_F)]$$

$$+ \frac{1}{\lambda(\theta_{FO}, y_F)} \left\{ \frac{y_h(\theta_{FO})}{c_h} + 1 - \frac{2[\beta_h c_h y_h(\theta_{FO})]^2}{\lambda^2(\theta_{FO}, y_F)} \right\} I_1[\lambda(\theta_{FO}, y_F)], \qquad (21)$$

where $I_1$ is the 1st order modified Bessel function of the first kind.

(2) The invariant multiplicity distributions of leading particles

It is believed that the leading particles are formed outside the nucleus, that is, outside the colliding region [36, 37]. The generation of leading particles is therefore free from fluid evolution. Hence, their rapidity distributions are beyond the scope of hydrodynamic description and should be treated separately.

In our previous work [24-26], we once argued that the rapidity distributions of leading particles take the Gaussian form

$$\frac{dN_{Lead}(b, \sqrt{s_{NN}}, y)}{dy} = \frac{N_{Lead}(b, \sqrt{s_{NN}})}{\sqrt{2\pi}\sigma} \exp\left\{ -\frac{[|y| - y_0(b, \sqrt{s_{NN}})]^2}{2\sigma^2} \right\}, \qquad (22)$$

where $y_0(b, \sqrt{s_{NN}})$ is the central position of distributions, which should increase with incident energies and centrality cuts. $\sigma$ in Eq. (22) is the width of distributions, which should not, at least not apparently, depend on the incident energies, centrality cuts and even colliding systems. Both $y_0$ and $\sigma$ can be determined by tuning the theoretical predictions to experimental data. $N_{Lead}(b, \sqrt{s_{NN}})$ in Eq. (22) is the number of leading particles, which, for an identical nucleus-nucleus collision, equals half of the number of participants.

The investigations have shown that [38], for certain rapidity, the invariant multiplicity distributions of leading particles possess the form

$$\frac{d^2 N_{lead}}{2\pi p_T dy dp_T} \propto \exp(-a p_T^2), \qquad (23)$$

where $a$ is a constant. Then, as a function of rapidity, the invariant multiplicity distributions of leading particles can be written as

$$\frac{\mathrm{d}^2 N_{\text{lead}}}{2\pi p_T \mathrm{d}y \mathrm{d}p_T} = \frac{\mathrm{d}N_{\text{Lead}}}{\mathrm{d}y} \frac{a}{\pi} \exp(-a p_T^2), \qquad (24)$$

which is normalized to $N_{\text{lead}}$.

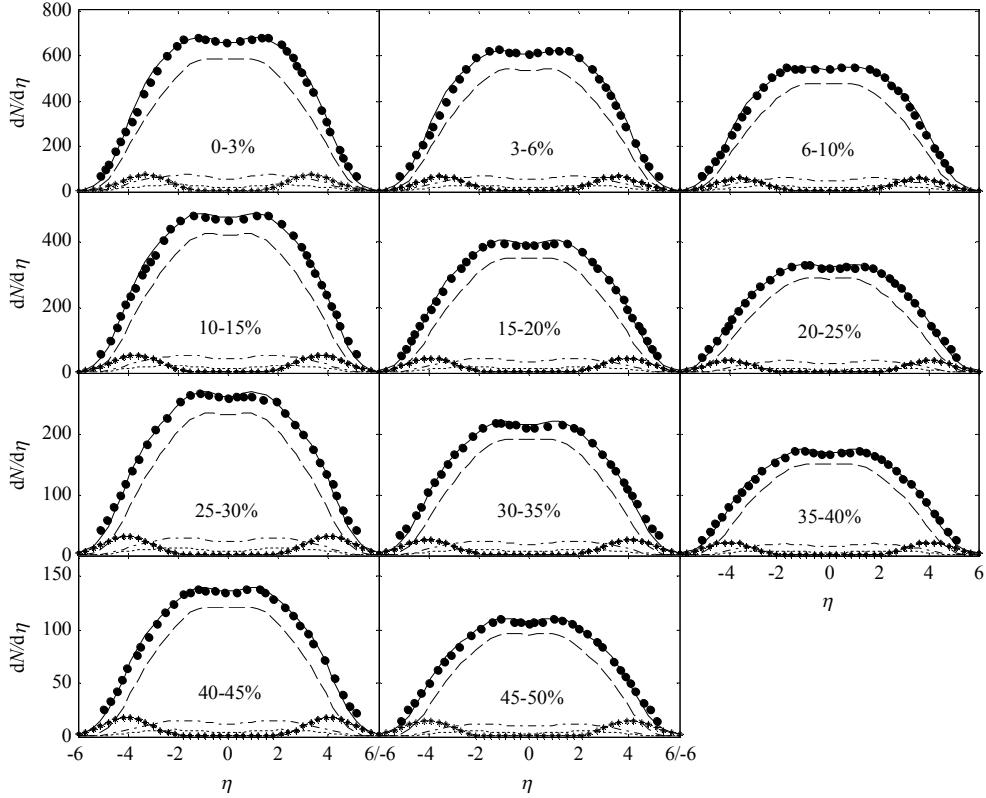

FIGURE 1: The pseudorapidity distributions of charged particles produced in different centrality Au+Au collisions at $\sqrt{s_{\text{NN}}} = 200$ GeV. The dashed, dashed-dotted and dotted curves are respectively the contribution from pions, kaons and protons got from the hydrodynamic result of Eq. (18). The dotted-star curves are the components of leading particles obtained from Eq. (24). The solid curves are the sums of the four types of curves.

(3) The pseudorapidity distributions of charged particles

Writing invariant multiplicity distributions in terms of pseudorapidity, we have

$$\frac{\mathrm{d}^2 N}{2\pi p_T \mathrm{d}\eta \mathrm{d}p_T} = \sqrt{1 - \frac{m^2}{m_T^2 \cosh^2 y}} \frac{\mathrm{d}^2 N}{2\pi p_T \mathrm{d}y \mathrm{d}p_T}, \qquad (25)$$

where

$$\frac{\mathrm{d}^2 N}{2\pi p_T \mathrm{d}y \mathrm{d}p_T} = \frac{\mathrm{d}^2 N_{\text{sQGP}}}{2\pi p_T \mathrm{d}y \mathrm{d}p_T} + \frac{\mathrm{d}^2 N_{\text{lead}}}{2\pi p_T \mathrm{d}y \mathrm{d}p_T}. \qquad (26)$$

To fulfill the transformation of Eq. (25), another relation

$$y = \frac{1}{2} \ln \left[ \frac{\sqrt{p_T^2 \cosh^2 \eta + m^2} + p_T \sinh \eta}{\sqrt{p_T^2 \cosh^2 \eta + m^2} - p_T \sinh \eta} \right] \qquad (27)$$

is in order.

Substituting Eqs. (18) and (24) into Eq.(25) and carrying out the integration of $p_T$, we can get the pseudorapidity distributions of charged particles produced in high energy heavy ion collisions. Figure 1 shows such distributions in Au+Au collisions at $\sqrt{s_{NN}} = 200$ GeV. The solid dots in the figure are the experimental measurements [5]. The dashed, dashed-dotted and dotted curves are respectively the contribution from pions, kaons and protons got from the hydrodynamic result of Eq. (18). The dotted-star curves are the components of leading particles obtained from Eq. (24). The solid curves are the sums of the four types of curves. The $\chi^2/\text{NDF}$ for each curve is listed in Table 1. It can be seen that the combined contribution from both hydrodynamics and leading particles matches up well with experimental data.

Table 1: The $\chi^2/\text{NDF}$, initial temperature $T_0$ and central position $y_0$ in different centrality Au+Au collisions at $\sqrt{s_{NN}} = 200$ and 19.6 GeV, respectively.

| Centrality cuts (%) | | 0-3 | 3-6 | 6-10 | 10-15 | 15-20 | 20-25 | 25-30 | 30-35 | 35-40 | 40-45 | 45-50 |
|---|---|---|---|---|---|---|---|---|---|---|---|---|
| $\chi^2/\text{NDF}$ | 200 GeV | 0.496 | 0.693 | 0.402 | 0.260 | 0.310 | 0.217 | 0.354 | 0.151 | 0.161 | 0.123 | 0.089 |
| | 19.6 GeV | 0.816 | 0.533 | 0.359 | 0.614 | 0.802 | 0.309 | 0.559 | 0.496 | 0.229 | 0.577 | --- |
| $T_0$(GeV) | 200 GeV | 0.950 | 0.949 | 0.948 | 0.947 | 0.945 | 0.941 | 0.922 | 0.909 | 0.885 | 0.874 | 0.860 |
| | 19.6 GeV | 0.551 | 0.549 | 0.547 | 0.543 | 0.539 | 0.533 | 0.529 | 0.518 | 0.510 | 0.493 | --- |
| $y_0$ | | 2.86 | 3.03 | 3.13 | 3.23 | 3.46 | 3.54 | 3.55 | 3.57 | 3.58 | 3.59 | 3.60 |

Experiments have shown that the overwhelming majority of charged particles produced in Au+Au collisions at $\sqrt{s_{NN}} = 200$ GeV consists of pions, kaons and protons with proportions of about 84%, 12% and 4%, respectively [39], which are roughly independent of energies, centrality cuts and colliding systems. In calculations, the ratios of these three kinds of particles take about the same as these values. $T_c$ in Eq. (15) takes the well-recognized value of $T_c = 180$ MeV. $c_h$ in Eq. (15) takes the value of $c_h = 0.45$ from the investigations of Refs. [15, 40-42]. The freeze-out temperature $T_{FO}$ takes the values of $T_{FO} = 120$ MeV from the studies of Ref. [6], which also shows that the baryochemical potential $\mu_B$ in Eq. (18) is about equal to 20 MeV. For the most central collisions, $T_0$ in Eq. (15) takes the value of $T_0 = 0.95$ referring to that given in Ref. [15]. This allows us to determine the constant $q_0$ in Eq. (16) to be $q_0 = 7.38 \times 10^{-4}$, $6.01 \times 10^{-4}$ and $4.50 \times 10^{-3}$ for pions, kaons and protons, respectively. Keeping $q_0$ unchanged, $T_0$ is fixed for the rest centrality cuts by making theoretical results fit in with experimental data. The results are listed in Table 1. It can be seen that $T_0$ decreases slowly with increasing centralities especially in the first four cuts. Table 1 also lists the central position $y_0$ in Eq. (22). As addressed above, it increases with increasing centralities. The width parameter $\sigma$ in Eq. (22) values a constant of $\sigma = 0.90$, being independent of centrality cuts. The parameter $a$ in Eq. (24) takes the value of $a = 0.92$.

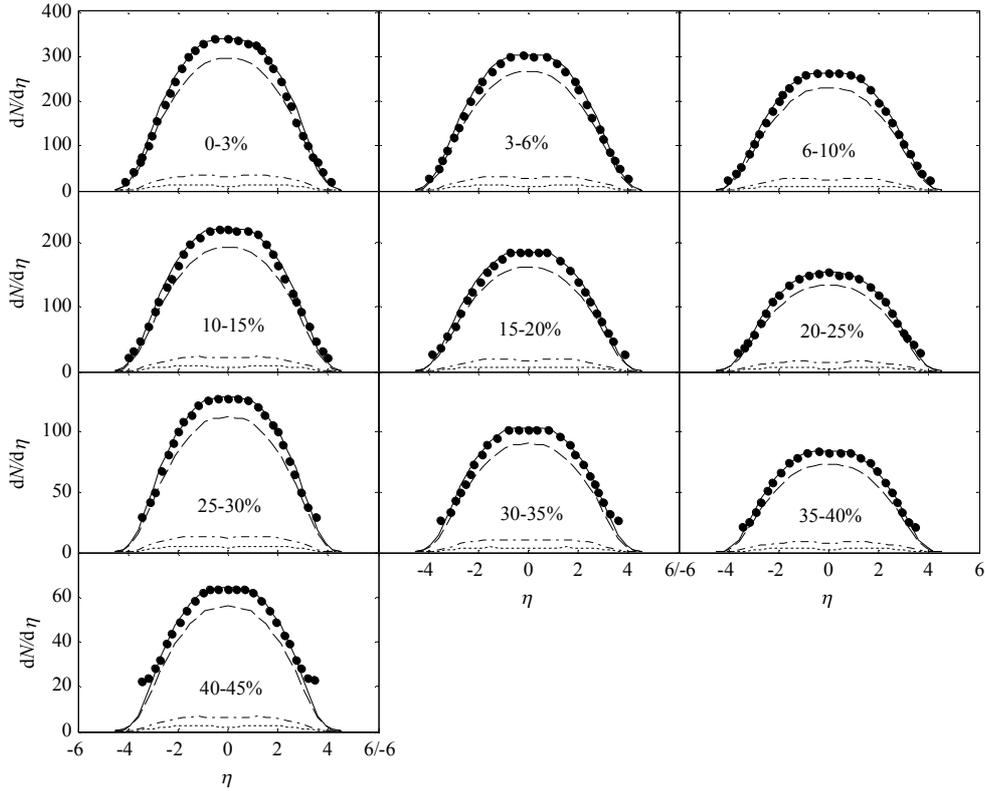

FIGURE 2: The pseudorapidity distributions of charged particles produced in different centrality Au+Au collisions at $\sqrt{s_{NN}} = 19.6\,\text{GeV}$. The dashed, dashed-dotted and dotted curves are respectively the contribution from pions, kaons and protons got from the hydrodynamic results of Eq. (18). The solid curves are the sums of the three types of curves.

Figure 2 shows the pseudorapidity distributions of charged particles produced in Au+Au collisions at $\sqrt{s_{NN}} = 19.6$ GeV. The $\chi^2/\text{NDF}$ for each curve is listed in Table 1. The meanings of different types of curves are the same as those in Figure 1. It can be seen that, in the absence of leading particles, the hydrodynamics alone can give a good description to the experimental observations. This is different from Figure 1. Where, leading particles are essential in fitting experimental data. This difference might be caused by the different transparencies of nuclei in different energies. As the analyses given in Ref. [43], in central Au+Au collisions at $\sqrt{s_{NN}} = 200$ GeV, the leading particles locate at about $y_0 = 2.91$. This position is far away from the mid-rapidity region. Where, relative to the low yields of charged particles frozen out from sQGP, the effect of leading particles is evident which should be considered separately. On the contrary, in case of Au+Au collisions at $\sqrt{s_{NN}} = 19.6$ GeV, $y_0 = 1.28$. This position is so close to the mid-rapidity region that the effect of leading particles is hidden by the large yields of charged particles generated from the freeze-out of sQGP. Therefore, there is no need to consider the contribution of leading particles separately.

In drawing Figure 2, $T_0$ takes the values as those listed in Table 1. $c_h = 0.40$ and $\mu_B = 210$ MeV. The other parameters, such as $T_c$, $T_{FO}$ and $q_0$ are the same as those used in drawing Figure 1.

## 4. Conclusions

By taking into consideration the effect of leading particles, the hydrodynamic model incorporating the phase transition is used to analyze the pseudorapidity distributions of charged particles produced in Au+Au collisions at RHIC energies.

The hydrodynamic model contains a rich information about transport coefficients of sQGP, such as the sound speed $c_0$ in sQGP, the sound speed $c_h$ in hadronic phase, the phase transition temperature $T_c$, the chemical freeze-out temperature $T_{FO}$, the baryochemical potential $\mu_B$ and the initial temperature $T_0$. With the exception of $T_0$, the other five coefficients take the values either from the well-known theoretical results or from experimental measurements. As for $T_0$, there are no widely accepted results so far. In our calculations, $T_0$ in the most central Au+Au collisions at $\sqrt{s_{NN}} = 200$ GeV takes the value referring to that given by other investigations, which enables us to ascertain the constant $q_0$ in Eq. (16). In the rest centrality cuts and in Au+Au collisions at $\sqrt{s_{NN}} = 19.6$ GeV, $T_0$ is determined by maintaining $q_0$ unchanged and comparing the theoretical results with experimental data.

The leading particles, by conventional definition, are the particles carrying on the quantum numbers of colliding nucleons and taking away the most part of incident energy. They are separately in projectile and target fragmentation region. The present investigations show that the importance of leading particles in describing the pseudorapidity distributions of charged particles produced in heavy ion collisions is related to the incident energy. At high energy, owing to the high transparency of nuclei, the contribution of leading particles is evident and indispensable. While, at low energy, as a result of poor transparency of nuclei the effect of leading particles is integrated with the results of freeze-out of sQCD. It does not need to be dealt with separately.

## Conflict of Interests

The authors declare that there is no conflict of interests regarding the publication of this paper.

## Acknowledgments

This work is supported by the Shanghai Key Lab of Modern Optical System.